\documentclass[aps,prl,twocolumn,groupedaddress,floatfix]{revtex4-1}

\usepackage{graphicx}

\usepackage{color}

\begin{document}


\title{Quantum quenches and off-equilibrium dynamical transition in the infinite-dimensional Bose-Hubbard model}


\author{Bruno Sciolla and Giulio Biroli}
\affiliation{Institut de Physique Th\'eorique, CEA/DSM/IPhT-CNRS/URA 2306 CEA-Saclay,
F-91191 Gif-sur-Yvette, France}


\date{\today}

\begin{abstract}
We study the off-equilibrium dynamics of the infinite dimensional Bose Hubbard Model
after a quantum quench. The dynamics can be analyzed exactly by mapping it to an
effective Newtonian evolution. For integer filling, we find a dynamical transition separating
regimes of small and large quantum quenches starting from the superfluid state.
This transition is very similar to the one found for the fermionic Hubbard model by mean field approximations.
\end{abstract}


\maketitle
Significant advances in the field of ultra cold atoms have allowed one to engineer quantum many-body systems in almost perfect isolation from the environment. Thanks to the ability to rapidly tune different parameters, e.g. the interaction strength between the atoms or the creation of controlled excitations, the realm of non-equilibrium many body physics of (almost) isolated quantum systems has thus been accessed and can now be studied experimentally. For example, Greiner et al. \cite{greiner} studied the dynamics of interacting bosons loaded on an optical lattice. The physics of this system is well captured
by the Bose-Hubbard model. By changing the intensity of the lasers one can effectively tune the parameters in the corresponding 
Bose-Hubbard model. Rapid changes induce interesting non-equilibrium dynamics \cite{greiner}. The activity in this field is booming: several new experiments have been performed,
including on fermionic systems \cite{greiner2,esslinger}; questions about thermalization \cite{rigol,biroli}, its absence \cite{weiss,kollath,roux}, quantum dynamical phase transitions out of equilibrium \cite{werner,schiro} are currently addressed.\\
A protocol inducing an off-equilibrium dynamics, which has received a lot of attention recently, is the so called
quantum quench. It corresponds to preparing the system in the ground state of the Hamiltonian $\hat{H}_i$, to changing suddenly 
at time $t=0$ a parameter of the Hamiltonian, for example the interaction strength, and then letting the system evolve 
with the new Hamiltonian $\hat{H}_f$. Several studies have been performed for the Bose Hubbard model, which as discussed
above is relevant for experiments. There have been numerical analysis of one dimensional systems by exact diagonalization and t-DMRG \cite{rigol,kollath,biroli,roux}.  Saddle point approximation \cite{altman}, Gross-Pitaevskii equations \cite{sachdev} and Gutzwiller approximation \cite{gutzb1,gutzb2} have been used to analyze higher dimensional and realistic cases. 
The fermionic Hubbard model has been also studied by mean field theories recently \cite{werner,schiro}.\\
In this work we present a complete analysis of quantum quenches in the Bose Hubbard model (BHM) in the limit of infinite 
dimensions. The advantage of this limit is that the model can then be analyzed exactly even out of equilibrium. 
Its solution at equilibrium played an important role in determining the phase diagram and the properties of the 
Mott-superfluid quantum phase transition of the three dimensional BHM \cite{sachdevbook}. Studying its off-equilibrium
dynamics is therefore a natural route to follow. We will discuss in the conclusion the limitations of this approach and possible
extensions.  To obtain a well defined infinite dimensional limit one has to scale the hopping amplitude as 
one over the dimension $d$\cite{grinstein}.  A complementary but in the bosonic case identical procedure \footnote{It can be shown adapting to the bosonic case the proof of A. Georges and J.S. Yedidia, J. Phys. A {\bf 24} (2173) 1991.}, which we will follow 
for simplicity, consists in focusing from the start on the BHM defined on a completely connected lattice. The corresponding Hamiltonian reads:
\begin{equation}
\label{Hamiltonian}
 \hat{H} = -\frac{J}{V} \sum_{i \neq j} \hat{b}^\dagger_j \hat{b}_i+\frac{U}{2}\sum_i \hat{n}_i(\hat{n}_i-1)
\end{equation}
where $\hat{b}^\dagger_i$, $\hat{b}_i$ are the bosonic creation and annihilation operators, $\hat{n}_i = \hat{b}^\dagger_i \hat{b}_i$ the occupation
operator and $V$ the total
number of sites. In the following we take $J=1$ and measure $U$ in units of $J$ and the time in units of $J/\hbar$. 
We shall study off-equilibrium dynamics induced by quantum quenches corresponding to a sudden change of 
the interaction strength from $U_i$ to $U_f$ at $t=0$.
Since $H$ is invariant under any permutation of sites, all eigenstates can be classified in terms of the corresponding symmetry classes. In particular the ground state, whether Mott or superfluid, corresponds to a completely site permutation symmetric wavefunction. Since also the time-dependent wavefunction remains completely symmetric after the quench, 
one can restrict the analysis to the subspace of completely symmetric states. It is easy to convince oneself that these states can be parametrized by the fraction $x_0,x_1,x_2,\ldots$ of sites with $0,1,2,\ldots$ bosons and that they correspond to the flat normalized sum of all Fock states characterized by $Vx_i$ sites with $i$ bosons per site. In order to simplify the presentation, let us first focus on the simplified model where maximum two bosons per site are allowed, $n_b=2$. 
We shall discuss later the generalization to any value of $n_b$. 
Since $x_0+x_1+x_2=1$ for $n_b=2$ and because the number of particles $V(x_1+2x_2)$ is conserved by the dynamics, a generic symmetric state is identified by $x_1$ only, and can be denoted $|x_1\rangle$ (henceforth we will drop the subindex $1$).
The evolution of the wavefunction {$|\psi \rangle = \sum_x \psi_x |x\rangle$} is determined by the equation $\langle x|i\partial_t \sum_{x'} \psi_{x'} | x' \rangle = \langle x| \hat{H} \sum_{x'} \psi_{x'} |x' \rangle$. In this model, all matrix elements $\langle x| \hat{H}  |x' \rangle$ are zero except $\langle x |\hat{H}|x \pm 2/V\rangle$ and the diagonal term $\langle x |\hat{H}|x \rangle$; the former corresponds to the physical process of one boson jumping from one site to another. The resulting Schr\"odinger equation for $\psi_x$ reads :
 \begin{equation}
\label{schrodinger}
\begin{array}{lll}
\displaystyle \frac 1 V i \partial_t \psi_{x}& = &
  \, D({x}) \psi_{x} -  W({x}) \Bigl(\psi_{x+2/V} + \psi_{x-2/V} \Bigr) \\
&= &\,\Bigl( D({x}) -  \, 2W({x}) \cosh (2\partial _{x}/V)\Bigr)\psi_{x} \\
&= &\,\Bigl( D({x}) -  \, 2W({x}) \cos (2\hat{p})\Bigr)\psi_{x} 
\end{array}
\end{equation}
\begin{figure}  
\begin{centering}
 \includegraphics[height=4.cm, bb = 0 0 205 138]{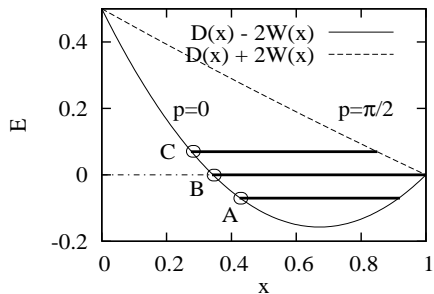}
  \caption{\label{fig1} Graphical solution for the value of $p$ at the turning points. The trajectories are full lines, and the position at $t=0$ is indicated by a circle for the three trajectories A, B and C. In case A it is impossible to have a turning point at $p=\pi/2$. Case B corresponds to the dynamical transition, and C to unbounded evolution of $p$. A, B and C are plotted in Fig. \ref{fig2}.}
  \end{centering}
\end{figure}
where $W(x)= x [(2-x-n)(n-x)/2]^{1/2}$, $D(x)= U (n-x)/2 -  \,  x(2+n-3x)/2$, $n$ is the number of 
bosons per site and subleading contributions in $1/V$ have been dropped. The initial wavefunction, which is the ground state at coupling $U_i$, is a wave packet of width $1/\sqrt{V}$, see \cite{epaps} and below. Since $1/V$ plays the role of $\hbar$ in (\ref{schrodinger}) the thermodynamic
limit corresponds to the \emph{classical} limit.{ As a consequence, the time evolution of the 
average particle position $x(t) = \langle \hat{x}\rangle$ and momentum $p(t) = \langle \hat{p} \rangle=\langle -i\partial_x/V\rangle$, is given by the Newton equations for the Hamiltonian $H=D(x)-2W(x)\cos(2p)$, where 
$x(t)$ and $p(t)$ are classical canonical variables. The validity of this argument can be thoroughly established by a direct analysis \cite{epaps}. In particular, one can show that  on timescales less than $\sqrt{V}$, $\psi_x(t) \sim \exp[V(x-x(t))^2/2\sigma(t)^2 +iVp(t)x]$, i.e. it is a sharp wave-packet, centered at $x(t)$, of width of order $1/\sqrt{V}$ and has a very fast oscillating phase $e^{iVp(t)x}$, where $x(t)$ and $p(t)$ are the classical canonical variables defined above.} 
In the following we will repeatedly make use of
this mapping to a classical system. Similar mappings have been recently used in \cite{gurarie,keeling,schiro}. 
The first useful consequence is that the ground state is obtained minimizing
 $H$ with respect to $p$ and $x$ ; the corresponding $p$ is actually always zero; in consequence the ground state is obtained by the value of $x$ minimizing $D(x)-2W(x)$.\\
The phase diagram is similar to the one derived by Fisher et al. \cite{grinstein} except that there is only
one Mott lobe corresponding to $n=1$. As we shall see, it is at integer filling, $n=1$, where the Mott state exists, 
that the off-equilibrium dynamics is most interesting. We shall consider this case first, for which
the ground state (GS) corresponds to
\[
x_{\mbox{\tiny GS}} = \left \{
\begin{array}{lll}
\displaystyle 1 & \textrm{ if } U \geq U_c, & \textrm{Mott insulator GS}\\
\displaystyle (U/U_c +1)/2 & \textrm{ if } U<U_c, & \textrm{Superfluid GS}\\
\end{array} \right .
\]
with $U_c=3+2\sqrt 2$.
\begin{figure}
 \includegraphics[height=4.cm,clip]{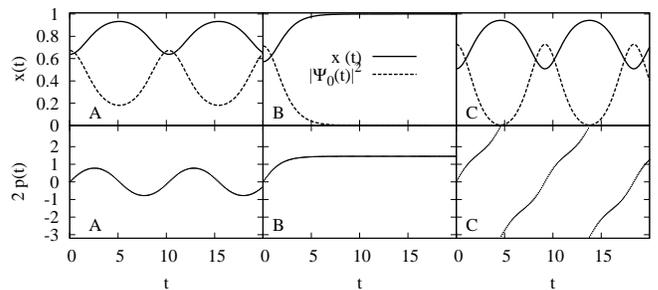}
 \caption{\label{fig2} Evolution as a function of time of $x,|\Psi_0|^2$ (top panels) and $p$ (bottom panels) increasing the amplitude of the quench. $U_f=3.33$ is kept fixed and several $U_i$ are considered, with A, B, C corresponding to $U_i=\{ 1.62,0.838,0.1 \}$ (unlike in the text where $U_f$ is varied). The case B correspond to the transition point.}
 \end{figure}
In this simple model, the condensate fraction $|\Psi_0|^2$ is simply equal to $\frac{1}{V^2} \sum_{i \neq j} b^\dagger_j b_i$, which up to a sign coincides with the average value of the (intensive) kinetic energy. This can be easily obtained subtracting  the average value of the interaction term to the total energy and reads, for the ground state, $|\Psi_0|^2=x_{\mbox{\tiny GS}}(1-x_{\mbox{\tiny GS}})U_c/2$. \\
Let us now consider quenches starting from a superfluid ground state
and increasing the value of $U$ from $U_i$ to $U_f$. A small increase of $U$ leads to oscillations of $x$ and $p$ as can be verified analytically and checked numerically, see Fig \ref{fig2}A. The turning points of $x(t)$, determined by $\dot{x}=4W(x)\sin(2p)=0$,  correspond to $p=0$. Actually there would be the possibility to have $p=n\pi/2$ too. However, 
a $p$ starting from zero and reaching the value $\pi/2$ would imply, by energy conservation, a value of $x$ at the turning point such that $E=D(x)+2W(x)$, where $E$ is the energy after the quench. 
This equation has no solution for small quenches as shown graphically in Fig. \ref{fig1}, see case A. 
It starts to have a solution for larger quenches, when $E$ becomes positive (cases B and C in Fig. \ref{fig1}). Actually $E_d=0$ corresponds to a \emph{dynamical transition}: for $E<E_d$ the momentum $p(t)$ is bounded, whereas for  $E>E_d$ it
grows infinitely large. The condition $E=0$ depends on $U_i,U_f$. One finds that for a given $U_i$, the corresponding critical value is $U_f^d(U_i)=(U_i+U_c)/2$. Approaching $U_f^d$ the period of oscillation increases and diverges
as $\tau=-c^{-1}\ln (|U_f-U_f^d|)$, where $c=\sqrt{(U_c-U_f)(U_f-1/U_c)}$. Fig. \ref{fig2} shows the typical time-evolution of $x,|\Psi_0|^2$ and $p$ for the three cases A, B and C.
Exactly at $U_f^d$ the system relaxes exponentially to the Mott state with a rate $c$. 
Approaching the transition the system spends most of the time close to the Mott state and therefore the time averaged condensate fraction vanishes at $U_f^d$ in a singular way, proportional to $1/\tau$. 
This singularity is related to the fact that the Mott state is `absorbing': classical trajectories falling into it cannot escape, and the period $\tau$ diverges when approaching $U_f^d$. { Conversely, trajectories starting from the Mott 
state remain stuck to $x=1$ on large times, $t \sim \log V$. This is, however, a peculiarity of the infinite dimensional limit; for a finite dimensional system, spatial fluctuations will drive the system away from the Mott state \cite{Zurek, Fischer}.}
In Fig. \ref{fig3}a, as an example of singular behavior, we show $|\Psi_0|^2$ as a function of $U_f$ for quenches starting from the non interacting case $U_i=0$. 
{ Moreover, we compare $|\Psi_0|^2$ to its microcanonical average at the same energy. Clearly, the system is not thermalized. At $U_f^d$ the condensate fraction goes to zero after the quench, whereas the corresponding equilibrium state is still superfluid.} 
The  dynamical phase diagram in Fig. \ref{fig3}b summarizes our analysis for all kinds of quantum quenches \footnote{Note that there is an additional dynamical transition obtained when quenching from the superfluid to very small $U$ ($U_f<1/U_c$) that we do not show since it is not robust: 
it disappears when one focuses on maximum three bosons per site ($n_b=3$). Details will be shown in \cite{long}.}. 
Let us finally address the changes in the dynamical behavior when one quenches for non integer filling. 
 Since the `absorbing' Mott state disappears for  $n\neq 1$, it is natural to expect, as indeed we find, that going away from $n=1$ the dynamical transition disappears too, and transmutes into a cross-over that becomes more and more sharp approaching integer filling. Overall, the resulting physical picture is extremely similar to the one obtained recently for 
 the fermionic Hubbard model by a time dependent Gutzwiller approximation \cite{schiro}. \\
\begin{figure}
 \includegraphics[width=8.0cm,clip,bb = 50 50 330 167]{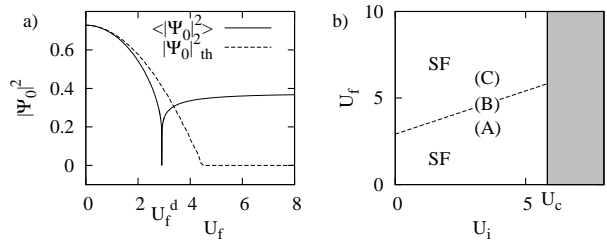}
 \caption{\label{fig3} a) Evolution of time average (continuous line) and microcanonical average (dashed line) of $\langle |\Psi_0|^2\rangle$ as a function of $U_f$ for $U_i=0$. b)  Dynamical phase diagram for the model with maximum two bosons per site. Quenches from the Mott phase are
not considered. Quenches from the superfluid phase are oscillating and similar to A or C. The dynamical transition separating the two is displayed as
a dashed line, it meets the Mott phase at $U_f = U_c$. The phase diagram for the case of more than two boson per site is qualitatively similar.}
 \end{figure} 
 Clearly, a natural question is how much these results depend on the constraint of maximum two bosons per site. { A complete analysis with an arbitrary number of bosons $n_b$ is very involved. The mapping to a classical system works
 also in these case. The classical degrees of freedom are the first $n_b-1$ fractions $x_0,x_1,\ldots,x_{n_b-1}$ and their associated canonical momenta.
Unlike in the case $n_b = 2$ where the classical motion is one dimensional, these trajectories are no longer necessarily periodic. In order to study 
their regularity we have computed numerically for $n_b = 3$ the largest Lyapunov exponent $\lambda$ \cite{lyapunov} of several trajectories. In this case $x_1,x_2$ are the classical variables and the expression of the Hamiltonian can be found in \cite{epaps}.} Depending on the initial condition we find large values ($\lambda>0.1$) characteristic of chaotic trajectories for large quenches, and small, possibly zero, values
characteristic of periodic or quasi periodic trajectories for small quenches.
{We find again a dynamical transition, for $n=1$ and $n=2$, which are the only filling for which the Mott ground state exists. At the transition line, the trajectories are chaotic.} As in the previous case, the dynamical transition corresponds to a change in the form of the phase space trajectories: for $U_f>U_f^d$ the momentum $2p_1-p_2$ becomes unbounded, see Fig. \ref{fig4}a. 
The time evolution of the $x_i(t)$ is also similar and characterized by oscillations that take place on longer timescales close to the transition. Moreover, the qualitative evolution of the time averaged $|\Psi_0|^2$ (and also $x_i$) with $U_f$ for a given $U_i$ resembles 
very much the one for $n_b=2$. We have also analyzed higher values of $n_b$ up to $n_b=5$ finding qualitatively and quantitatively similar results. Actually, the evolution of  $\langle |\Psi_0|^2 \rangle$ depends very little on $n_b$ for $n_b>2$ as shown in Fig. \ref{fig4}b (the two curves $n_b = 4$ and $n_b=5$ differ by less than $0.01\%$).
  \begin{figure}
  \begin{center}
    \includegraphics[height=4.cm,bb = 52 52 435 224]{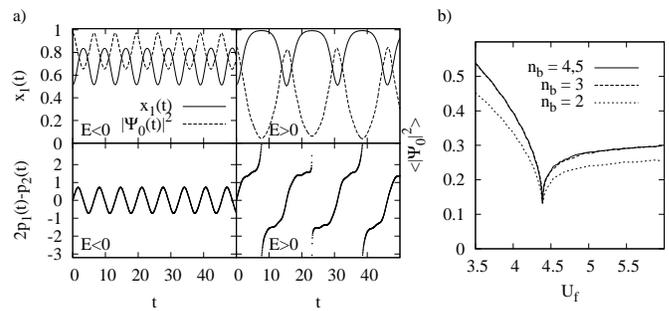}
   \end{center}
 \caption{\label{fig4}
 a) As Fig. \ref{fig2} but for $n_b = 3$ and $U_i=1$. Left and right: $U_f=2.5$ and $3.29$, respectively below ($E<0$) and above ($E>0$) the dynamical transition at $U_f^d = 3.21$. b) Variation of $\langle |\Psi_0|^2 \rangle$ as a function of $U_f$ for $n=1$, $U_i=3$, with $n_b = 2,3,4$ and $5$. The plot of $n_b = 2$ is shifted of $0.025$ along the $U_f$ axis for comparison.}
 \end{figure}
The only issue that remains open is the form of the singularity at the dynamical transition for $n_b>2$. 
Numerical solutions of the Newton equation are not precise enough to answer this question. Even in the case
of two bosons per site, for which we know that $|\Psi_0|^2=0$ at the transition and the singularity is logarithmic, 
numerics alone would not be conclusive. For $n_b=2$ the singularity was due to the fact that trajectories spend
most of the time close to the Mott state. For $n_b>2$, it is not clear whether trajectories starting with the same energy as the Mott state (energy zero) have to go arbitrary close to it. Assuming that the classical dynamics is completely ergodic on the $E=0$ hypersurface, one could argue that this should be the case. However, even in this case, time averages would not coincide with averages in the Mott state unless the recurrence time is of the same order as the trapping time, a difficult question to address.  
The conclusion of the analysis performed for higher number of bosons is that the results for $n_b=2$ are robust and expected to hold also for the BHM with an arbitrary number of bosons per site, except possibly the form of the singularity of $|\Psi_0|^2$ (and of the other observables).\\
Let us now discuss the implications and the limitations of our findings. Clearly, the infinite dimensional limit neglects important
dynamical and spatial fluctuations. This is manifest from the non damped oscillatory evolution in time of the observables and the absence
of thermalization. Certainly $1/d$ corrections must be taken into account to lead to 
decoherence and thermalization. Nevertheless, we expect that our mean field approach should be able to qualitatively 
account for the short time dynamical behavior. As a consequence, the dynamical transition we find should transmute into a cross-over in the short time dynamics for finite dimensional systems. Indeed results obtained for the one dimensional BHM seem to be in agreement with our findings \cite{corinna2}. 
{ Moreover, we expect the dynamical transition we found to be quite general, at least within mean field treatments of the off-equilibrium dynamics. Actually, it is qualitatively identical to the one found for the fermionic Hubbard model within the Gutzwiller approximation \cite{schiro} and very similar to the one obtained by out of equilibrium dynamical mean field theory \cite{werner}, where some dynamical fluctuations are taken into account.}\\
Including spatial and dynamical fluctuations would allow one to go beyond our mean field treatment.
{ A good description of decoherence and thermalization for the BHM could be obtained in the future within a real time generalization of the  equilibrium bosonic DMFT \cite{BDMFT1,BDMFT2}. }
We would like to thank C. Kollath and M. Schir\`o for useful discussions. GB acknowledges 
partial financial support from ANR FAMOUS.

\end{document}